\documentclass[journal,comsoc]{IEEEtran}

\usepackage[T1]{fontenc}
\usepackage{caption}
\usepackage{CJKutf8}
\usepackage{epsfig}
\usepackage{multirow}
\usepackage{enumitem}
\usepackage{hyperref}
\setitemize{noitemsep,topsep=0pt,parsep=0pt,partopsep=0pt}
\usepackage[noend]{algpseudocode}
\usepackage{algorithmicx,algorithm}
\usepackage{enumerate} 
\usepackage{pgfplots}

\usepackage{graphicx}
\usepackage{pythonhighlight}
\usepackage[english]{babel}
\usepackage[T1]{fontenc}
\usepackage{float}
\usepackage{bm}
\usepackage{braket}
\usepackage{xcolor}
\usepackage{amsfonts}
\usepackage{comment}
\usepackage{dirtytalk}
\usepackage{hyperref}
\usepackage{url}
\usepackage{bbold}
\usepackage{cite}
\usepackage{realboxes}

\usepackage{enumitem}
\usepackage{algpseudocode}

\usepackage{listings}
\usepackage{xcolor}
\definecolor{bkgd}{RGB}{240,242,246}
\definecolor{ceruleanblue}{rgb}{0.16, 0.32, 0.75}
\definecolor{orange-red}{rgb}{1.0, 0.27, 0.0}
\definecolor{anotherblue}{RGB}{37,92,243}
\definecolor{blackblue}{RGB}{46,60,85}
\definecolor{goldyellow}{RGB}{199,146,12}

\usepackage{amssymb}
\usepackage{bbding}
\usepackage[caption=false]{subfig}
\usepackage{booktabs}

\usepackage{tabularx}

\usepackage{ragged2e}    

\usepackage[resetlabels]{multibib}
\usepackage{textcomp}

\usepackage{amsmath}
\usepackage{cleveref}
\usepackage[cmintegrals]{newtxmath}
\begin{document}

\begin{CJK*}{UTF8}{gbsn}

\title{Quantum-Enhanced LLM Efficient Fine Tuning}
\author{
  \IEEEauthorblockN{Xiaofei Kong,$^{1}$ 
  Lei Li,$^{1}$ 
  Zhaoyun Chen,$^{2}$ 
  Cheng Xue,$^{2}$
  Xiaofan Xu,$^{4}$
  Huanyu Liu,$^{4}$
  Yuchun Wu,$^{4}$ 
  Yuan Fang,$^{1}$
  Han Fang,$^{5}$
  Kejiang Chen,$^{6}$
  Yang Yang,$^{2, 7}$ 
  Menghan Dou,$^{1, 3}$ 
  Guoping Guo,$^{4,*}$\thanks{* Corresponding author: gpguo@ustc.edu.cn}}
  \\[2ex]
  
  \IEEEauthorblockA{
    $^{1}$Origin Quantum Computing Company Limited, Hefei 230088, China\\
    $^{2}$Institute of Artificial Intelligence, Hefei Comprehensive National Science Center, Hefei, Anhui 230088, China\\
    $^{3}$Anhui Engineering Research Center of Quantum Computing, Hefei 230026, China\\
    $^{4}$Laboratory of Quantum Information, University of Science and Technology of China, Hefei 230026, China\\
    $^{5}$National University of Singapore, Kent Ridge 119077, Republic of Singapore\\
    $^{6}$University of Science and Technology of China, Hefei 230026, China\\
    $^{7}$Anhui University, Hefei, 230039, China\\
  }
  \thanks{This work has been supported by the National Key Research and Development Program of China (Grant Nos.  2024YFB4504101 and 2023YFB4502500), and the Anhui Province Science and Technology Innovation (Grant No. 202423s06050001).}
}

\maketitle
\begin{abstract}

Low-Rank Adaptation (LoRA) enables efficient fine-tuning of pre-trained language models through low-rank matrix approximation, achieving effectiveness in many scenarios. However, its representation capacity is constrained in complex tasks or high-rank dependency settings, potentially limiting model adaptability. To overcome the expressive bottleneck in classical low-rank approximation for fine-tuning large language models (LLMs), we propose \textbf{Q}uantum \textbf{T}ensor \textbf{H}ybrid \textbf{A}daptation (QTHA), a parameter-efficient fine-tuning method that integrates a quantum neural network (QNN) with a tensor network. QTHA explores quantum tensor hybrid fine-tuning within low-rank spaces by decomposing pre-trained weights into quantum neural network and tensor network representations, leveraging quantum state superposition to overcome classical rank limitations. Experiments demonstrate that QTHA achieves performance comparable to or surpassing LoRA in parameter-efficient fine-tuning. Compared to LoRA, QTHA reduces trainable parameters by 76\% while reducing training loss by up to 17\% and improving test set performance by up to 17\% within the same training steps. This research not only enables lightweight adaptation of quantum resources to the billion-parameter models but also validates the feasibility of quantum hardware optimization driven by LLM tasks. It establishes the first engineering-ready foundation for future quantum-enhanced Artificial General Intelligence (AGI) systems.  

\end{abstract}
\IEEEpeerreviewmaketitle

\section{Introduction}

The rapid advancement of large language models (LLMs) has driven innovations in parameter-efficient fine-tuning (PEFT) to reduce computational overhead while preserving performance. Classical methods such as Low-Rank Adaptation (LoRA) \cite{hu2022lora} and Weighted-Decomposed Low-Rank Adaptation (DoRA) \cite{liu2024dora} assume that weight updates during fine-tuning lie within low-rank subspaces, enabling efficient adaptation via trainable low-rank matrices. Similarly, prefix-tuning optimizes task-specific vectors appended to model inputs, mimicking “virtual tokens” to guide downstream tasks without altering core parameters \cite{li-liang-2021-prefix}. While effective, these low-rank approximations inherently limit feature representation adaptability, impair convergence in complex tasks, and exhibit sensitivity to rank selection \cite{hu2022lora,liu2024dora,zhang2023adaptive}.

Recent quantum-inspired methodologies address these limitations through two complementary paradigms. The first leverages tensor-based adaptations, such as Quantum Tensor Adaptation (QuanTA) \cite{NEURIPS2024_a7c17115}, which employs quantum circuit-inspired tensor decomposition for high-order parameter adjustments, and Tensor Product Attention (TPA) \cite{zhang2025tensor}, which optimizes memory efficiency via contextual low-rank factorization. Parallel efforts, including CompactifAI \cite{tomut2024compactifai}, integrate tensor networks with singular value truncation for model compression. The second paradigm combines quantum neural network (QNN) \cite{sim2019expressibility} with classical architectures: Quantum-PEFT \cite{koike2024quantum} achieves logarithmic parameter scaling through entangled unitary transformations, while Quantum Parameter Adaptation (QPA) \cite{liu2025a} generates compact tuning parameters via hybrid quantum-classical mappings. These approaches integrate quantum-derived high-rank information into classical low-rank spaces, with Matrix Product Operator (MPO) representations \cite{gao2020compressing} further enhancing robustness through localized entanglement regularization.

Despite these advances, existing quantum-inspired frameworks remain largely theoretical, lacking validation on physical quantum hardware. To bridge this gap, we propose the Quantum Tensor Hybrid Adaptation (QTHA), which reparameterizes pre-trained layers into quantum tensor hybrid architectures. By synergizing QNN for capturing complex transformations with tensor-based efficiency, our framework achieves superior parameter-efficient fine-tuning while surpassing conventional LoRA in performance. Our main contributions include:

\textbf{First implementation of quantum computing inference for LLM on quantum hardware}:
We introduce QTHA, a novel quantum-enhanced fine-tuning algorithm for LLM, based on a hybrid quantum-classical neural network architecture. This framework synergistically combines the expressive power of QNN with the efficiency of tensor networks to achieve PEFT. Notably, QTHA represents the first practical implementation of inference technology on quantum hardware, bridging the gap between theoretical quantum machine learning and deployable solutions.

\textbf{Significant reduction in trainable parameters}:
The QTHA method demonstrates remarkable parameter efficiency, reducing trainable parameters by 76\% compared to LoRA under identical conditions while preserving model performance. This substantial parameter reduction significantly enhances model trainability. Notably, in specific scenarios, QTHA accelerates convergence by 20\% in terms of step size (measured by the rate of loss value change per unit time), enabling the model to achieve optimal fitting performance with fewer training iterations. This dual advantage not only reduces overfitting risks but also shows promising potential for lowering computational costs.

\textbf{Enhancing the performance of LLM fine-tuning}:
We demonstrate that QTHA not only serves as a viable alternative to classical parameter-efficient fine-tuning methods such as LoRA and DoRA, but also exhibits superior performance across a broader spectrum of fine-tuning tasks. Specifically, QTHA achieves up to a 17\% reduction in training loss while significantly decreasing the number of trainable parameters. Moreover, it improves accuracy metrics by up to 17\% on small-scale test sets through optimized parameter adaptation mechanisms.

This study employs a systematic framework to advance quantum-enhanced language model research: Section I establishes the innovative potential of this work in addressing domain-critical challenges by tracing the evolutionary trajectory of LLMs and synthesizing advancements in quantum-inspired adaptation methodologies. Section II deconstructs conventional fine-tuning paradigms while proposing a groundbreaking quantum-integrated strategy that architecturally embeds quantum computational principles into LLM adaptation frameworks. Section III designs an innovative hybrid experimental protocol, demonstrating statistically significant benchmarking results through comprehensive comparative analyses to elucidate the mechanism of resource-efficient quantum-enhanced fine-tuning. The concluding Section IV crystallizes dual-impact theoretical-practical innovations in quantum-enhanced fine-tuning of LLMs, charting transformative pathways for next-generation LLM ecosystems.

\section{Methods}

The QNN leverages the dynamic properties of quantum entanglement and superposition to achieve highly nonlinear feature modeling capabilities. In this framework, the MPO is responsible for efficiently extracting abrupt features, while the QNN focuses on effectively learning periodic features\cite{kordzanganeh2023parallel,gao2020compressing}. Through the linear combination of the two, this approach aims to overcome the limitations of classical linear layers in feature learning, thereby achieving optimal allocation of computational resources. Furthermore, quantum circuits can explore a broader solution space\cite{Farhi2014AQA}, circumventing the inherent limitations of local optima of classical low-rank models.

\subsection{Quantum Tensor Network Based on MPO}

To efficiently represent the low-rank weight matrix $ \mathbf{W} $ in LoRA, we employ a tensor decomposition method called the Matrix Product Operator (MPO) \cite{gao2020compressing}. By reorganizing the elements of matrix $ \mathbf{W} $ into a higher-dimensional tensor, we obtain a tensor with $ 2n $ indices.


Specifically, let the input space $ \mathcal{X} $ and output space $ \mathcal{Y} $ have dimensions $ N_x $ and $ N_y $, respectively. Through multilinear algebraic transformations, the original weight matrix $ \mathbf{W} \in \mathbb{R}^{N_y \times N_x} $ is mapped to a higher-order tensor:

\begin{equation}
    \mathcal{W} \in \mathbb{R}^{J_1 \times \cdots \times J_n \times I_1 \times \cdots \times I_n},
\end{equation}
where
\begin{equation}
    \prod_{k=1}^n I_k = N_x, \quad \prod_{k=1}^n J_k = N_y.
\end{equation}

Adopting a hierarchical index mapping strategy, we reshape the input vector $ x \in \mathcal{X} $ into a multidimensional tensor $ (i_1, i_2, \ldots, i_n) $, with the output vector corresponding to $ (j_1, j_2, \ldots, j_n) $. We then decompose the weight matrix $ \mathbf{W} $ into a product of local tensors $ {w}^{(k)} $ via tensor factorization:

\begin{equation}
\mathcal{W}_{j_1 \cdots j_n, i_1 \cdots i_n} = \mathrm{Tr}\left[ \mathbf{w}^{(1)}_{j_1 i_1} \mathbf{w}^{(2)}_{j_2 i_2} \cdots \mathbf{w}^{(n)}_{j_n i_n} \right],
\end{equation}
where in the local tensor $ \mathbf{w}^{(k)} \in \mathbb{R}^{D_{k-1} \times J_k \times I_k \times D_k} $ satisfies the following conditions：

\begin{equation}
    \mathbf{w}^{(k)}_{j_k i_k} \in \mathbb{R}^{D_{k-1} \times D_k}.
\end{equation}
Here, bond dimension parameter $ D=\max\{D_k\} $ governs the model's expressive capacity, with its value positively correlated to the quantum entanglement entropy. By establishing a controllable balancing mechanism between model complexity and expressive power, this approach provides a novel technical pathway for lightweight design of large-scale neural networks. For parameter optimization, the total number of trainable parameters satisfies:

\begin{equation}
    N_{\text{MPO}} = \sum_{k=1}^n I_k J_k D_{k-1} D_k,
\end{equation}
when a uniform bond dimension $ D_k \equiv D $ is adopted, and the total parameter count simplifies to:
\begin{equation}
    N_{\text{MPO}} = D(I_1 J_1 + I_n J_n) + D^2 \sum_{k=2}^{n-1} I_k J_k.
\end{equation}

Compared to the $ N_x \times N_y $ parameters of traditional fully connected layers, our method achieves exponential compression when $ n \geq 3 $. The initialization strategy employs an improved Kaiming uniform distribution~\cite{he2015delving}, this method of initialization effectively avoids parameter redundancy during the early stages of training while maintaining gradient stability.

\subsection{\textbf{Q}uantum \textbf{T}ensor \textbf{H}ybrid \textbf{A}daptation}

The core of quantum neural networks lies in their ability to leverage the superposition and entanglement of quantum states, enabling efficient representation of high-dimensional features in Hilbert space. We designed a highly expressive QNN, in which an input vector \textit{$ x \in \mathbb{R}^n $} is assigned to a quantum state \textit{$ |\psi(x)\rangle $} via angle embedding using $R_Y$ gates: 
\begin{equation}
|\psi(x)\rangle = \bigotimes_{i=1}^q R_Y(x_i) |0\rangle, 
\end{equation}
where $ q $ denotes the number of qubits.

The quantum state evolves through a parameterized unitary transformation $ U(\theta) $:  

\begin{equation}
|\phi(x)\rangle = U(\theta) |\psi(x)\rangle = \prod_{l=1}^L \left( \bigotimes_{i=1}^q R_Y(\theta_{l,i}) \cdot \text{CRZ}(\theta_{l,ij}) \right),
\end{equation}
where $ L $ is the number of layers. CRZ represents controlled-Z rotation gates, which introduces entanglement between qubits.  

Each qubit is measured through Pauli-Z observables, yielding expectation values:  
\begin{equation}
y_i = \langle \phi(x) | Z_i | \phi(x) \rangle \in [-1, 1],
\end{equation}
with the final output \textit{$ O_q = [y_1, \ldots, y_m] $} being a classical vector.  

By leveraging quantum state superposition and entanglement, QNN generate nonlinear features in high-dimensional Hilbert space. Although $ y $ is classical data, its generation process involves nonlinear transformations of quantum states, enabling the extraction of features that are difficult to capture with classical methods. While MPO neural networks can generate non-harmonic functions, QNN excels at fitting truncated Fourier series \cite{kordzanganeh2023parallel,PhysRevA.103.032430}, by leveraging quantum state superposition and entanglement, QNN generate nonlinear features in high-dimensional Hilbert space.

To address the limitations of classical neural networks in low-rank spaces, we linearly combine the classical output of the QNN, denoted as $ O_{q} $, with the output of the classical neural network, denoted as $ O_{c}$, $ W_q $ and $ W_c $ are the weights for the quantum neural network output and the classical neural network output, respectively, used to adjust the contributions of $ O_q $ and $ O_c $:

\begin{equation}
    \tilde{O} = W_q O_q + W_c O_c.
\end{equation}

The updated output $ \tilde{O} $ retains the low-rank feature learning capability while incorporating the high-dimensional features extracted by the QNN, thereby enhancing the model's ability to model complex nonlinear relationships. Through this linear operation, the QNN's output $ O_q $ provides additional nonlinear features to $ \tilde{O} $. The elements in $ \tilde{O} $ encompass both harmonic and non-harmonic features \cite{abbas2021power,zhou2023quantum,schuld2019quantum,biamonte2017quantum}, extracting harmonic features and complementing the classical neural network. As a result, the expressive power of the updated output $ \tilde{O} $ is theoretically enhanced.

\begin{figure}[! h]
\centering
\includegraphics[width=0.99\linewidth]{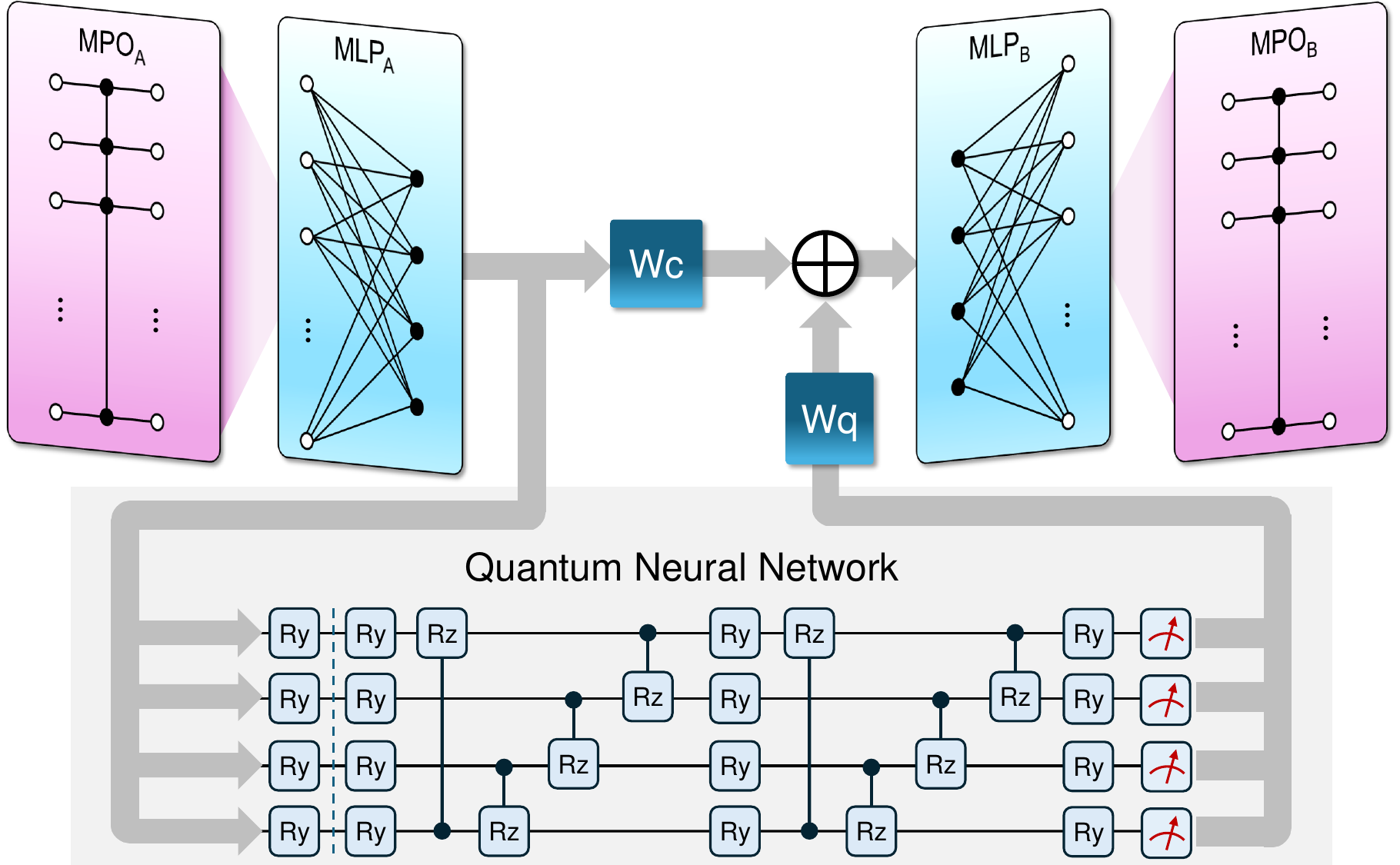}
\vspace{0.8em}
\caption{Schematic diagram of the Quantum Tensor Hybrid Adapation (QTHA). QTHA dynamically adjusts feature weights through parameter tuning and outputs a combination of features from MPO and QNN.}
\label{fig:QTHA}
\end{figure}

Based on the aforementioned MPO and quantum hybrid network, we construct the final QTHA, as shown in Fig.~\ref{fig:QTHA}. QTHA uses MPO to reduce the number of parameters in LoRA. Specifically, the input vector of LoRA is reshaped and fed into $\mathrm{MPO_A}$. The output from $\mathrm{MPO_A}$ is then processed by a classical multilayer perceptron ($\mathrm{MLP_A}$), which transforms it into a representation corresponding to the number of qubits in the subsequent QNN. Within the QNN, the input is first encoded using RY angles, followed by a variational quantum circuit. These operations leverage quantum superposition and entanglement to efficiently encode high-dimensional features. After performing Pauli-Z measurements on each qubit, a vector of the same length as the output of $\mathrm{MLP_A}$ is generated. The input for the next layer, $\mathrm{MLP_B}$, is computed as a weighted combination of the QNN output and the output from $\mathrm{MLP_A}$. Subsequently, $\mathrm{MLP_B}$ applies a linear transformation to this combined input, producing a vector that is then passed to $\mathrm{MPO_B}$ to generate the final output. All dimensions within this architecture can be configured as hyperparameters for flexibility and optimization. By using parameterized two-qubit gates CRZ (Controlled-Rotation-Z), we can adjust parameters to enhance the exploration of the state space. Our design adopts a block structure, where each block consists of continuous nearest-neighbor interactions and one non-local interaction. This modular approach supports performance enhancement through stacking multiple layers, yet achieves near-saturation of expressive power with fewer layers, thus reducing optimization complexity. 

\section{Experiment}

\subsection{Datasets}

The datasets used in this study are publicly available, specifically the CPsyCoun~\cite{zhang-etal-2024-cpsycoun} Chinese dataset, the R1-Distill-SFT~\cite{slam-distillation-from-r1} English dataset and the Chinese-DeepSeek-R1-Distill-data-110k (CH-R1-Math)~\cite{Chinese-Data-Distill-From-R1}.

CPsyCoun dataset is a Chinese psychological counseling dialogue dataset from HuggingFace (by CAS-SIAT-XHai), featuring anonymized, multi-turn dialogues (16K samples) across 9 common issues (e.g., depression, anxiety) and 7 counseling schools. Includes counselor-client role labels, some with fine-grained annotations (e.g., strategies, emotional support). Ideal for counseling dialogue generation, sentiment analysis, and intervention modeling in NLP for mental health.  

R1-Distill-SFT dataset is a K-12 math QA dataset with structured questions (arithmetic, algebra, geometry) and step-by-step solutions. Includes question stems, multilingual labels (e.g., English/Chinese), LaTeX-formatted answers, and verification status. Filtered for quality, it supports SFT, knowledge distillation, and automated evaluation in education.  
 
CH-R1-Math dataset is a Chinese instruction-tuning dataset (110K samples) from ModelScope, distilled from DeepSeek-R1. Covers multi-turn dialogues, QA, code generation, and math reasoning. Rigorously cleaned and deduplicated, it features diverse tasks, logical coherence, and contextual annotations. Optimized for fine-tuning models like DeepSeek-R1.

\subsection{Evaluation Metrics}

In this experiment, cross-entropy loss\cite{goodfellow2016deep} is adopted as the training and validation loss function for the model. Cross-entropy measures the difference between the predicted probability distribution of the model and the true label distribution, effectively reflecting the optimization level of classification tasks. A smaller value indicates that the model's predictions are closer to the true distribution. For a dataset containing $N$ samples, the cross-entropy loss is calculated as:

\begin{equation*}
\mathcal{L} = -\frac{1}{N}\sum_{i=1}^{N}\sum_{k=1}^{C} y_{ik} \log(p_{ik}),
\end{equation*}
where C represents the total number of classes in the classification task, $y_{ik}\in\{0,1\}$ is the true label of sample $i$ for class $k$ (in one-hot encoding form), and $p_{ik}\in(0,1)$ denotes the predicted probability that sample $i$ belongs to class $k$. This loss function averages the prediction errors across all samples through the double summation operation, effectively penalizing misclassified predictions while ensuring comparability of losses across different batches of data.

Perplexity (PPL) is a common evaluation metric in natural language processing used to measure how well a language model predicts a text sequence. It reflects the model's uncertainty when assigning probabilities to words in the test data. Lower perplexity indicates better performance. PPL is defined as:
$$
\text{PPL} = \exp\left(-\frac{1}{N} \sum_{i=1}^{N} \log P(w_i | w_{<i})\right),
$$
where $ N $ denotes the total number of words/tokens in the test data and $ P $ is the probability assigned by the model to the $ i $-th word given prior context. A lower PPL means that the model assigns higher probabilities to the correct words, showing better alignment with the true data distribution.

For the evaluation of text generation, we use BLEU-4~\cite{papineni2002bleu} and ROUGE~\cite{lin2004rouge}. BLEU-4 assesses lexical precision through 4-gram matching between generated and reference texts, while ROUGE measures content coverage and semantic coherence using recall-oriented n-gram and sequence alignment metrics. This combination provides complementary perspectives on the quality of the generation.

\subsection{Results}

Benchmark of various fine-tuning methods using DeepSeek-R1-Distill-Qwen-7B~\cite{deepseekai2025deepseekr1incentivizingreasoningcapability} and Qwen2-7B-Instruct~\cite{qwen2} models as the base model.The ranks of LoRA~\cite{hu2022lora} and QTHA are set to 4, and the decomposition factor of QuanTA~\cite{NEURIPS2024_a7c17115} is set to 5. QTHA, LoRA and QuanTA are applied to the linear projection layers q\_proj  and v\_proj.

\begin{figure*}[!t]
\centering
\includegraphics[width=\textwidth]{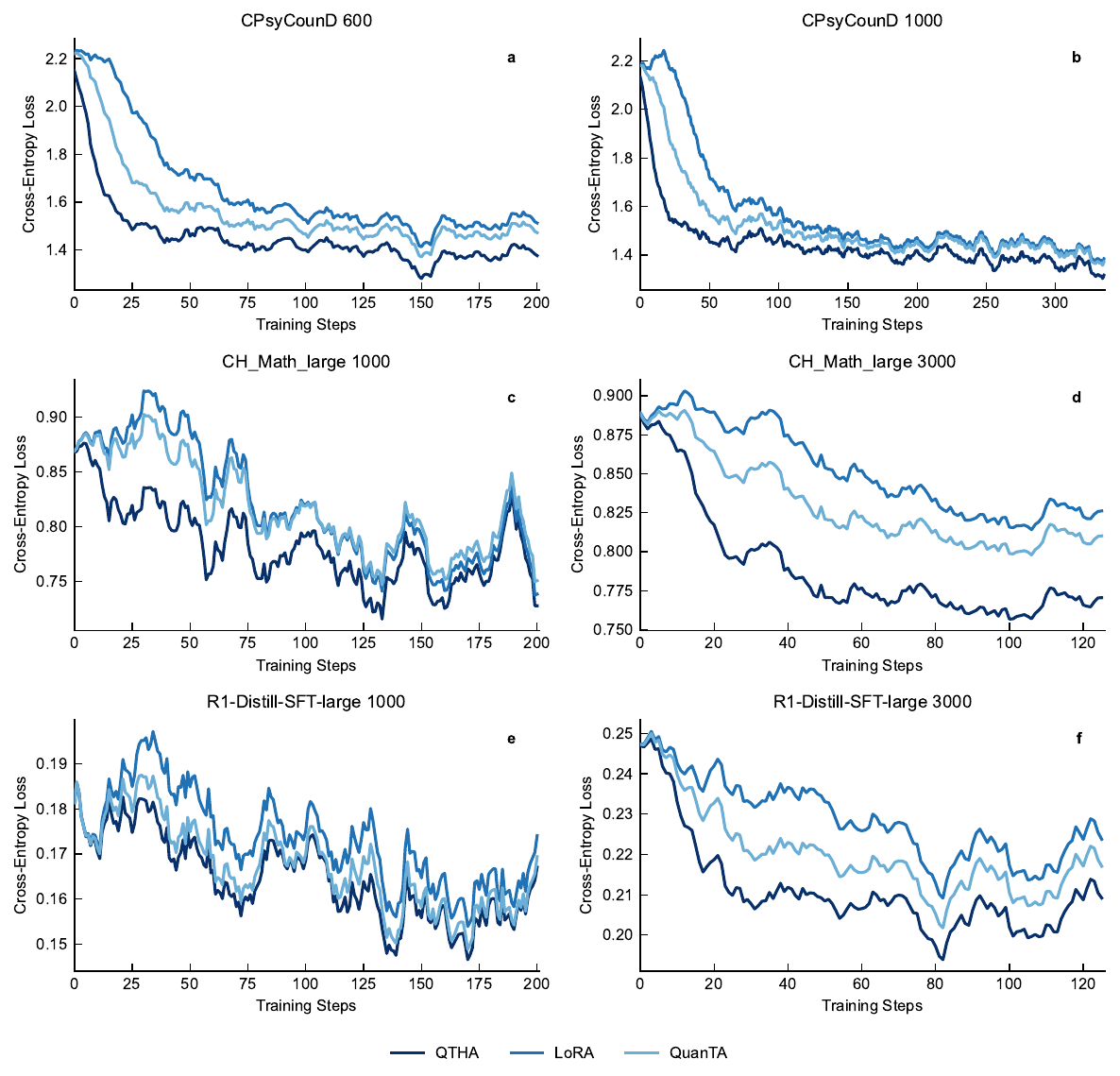}
\caption{Training loss comparison of QTHA, LoRA and  QuanTA across varying sample sizes (N) and datasets. Subplots (a–b) illustrate results on the CPsyCounD dataset, while subplots (c–d) and (e–f) show corresponding analyses on the R1-Distill-SFT and CH-R1-Math dataset. QTHA consistently achieves lower training losses compared to baseline methods, demonstrating enhanced convergence efficiency under diverse experimental conditions.}
\label{fig:train_result}
\end{figure*}

This section primarily aims to verify whether the efficient fine-tuning algorithm described for the QTHA can improve the performance of fine-tuned models while significantly reducing the number of parameters. The experiments were conducted using PyTorch~\cite{paszke2019pytorch} and PyVQNet~\cite{chen2019vqnet,bian2023vqnet} with quantum circuit simulations. As shown in Fig.~\ref{fig:train_result}, the convergence curve of QTHA exhibits a steeper descent rate during early training stages and achieves a lower stable loss value in training phase.

The datasets are extracted from CPsyCounD, R1-Distill-SFT, and CH-R1-Math, with 300, 600, 1000, 3000 respectively. The maximum sequence length max\_seq\_len for both training and prediction is set to 1024, with 10\% of the data used as the validation set and the remaining 90\% as the training set. For the test set, 30 to 100 samples from the aforementioned three datasets are extracted as the test set.

DeepSeek-R1-Distill-Qwen-7B exhibits excellent performance in code and mathematical reasoning tasks. we use the following metrics to perform a small-scale evaluation of the performance of the test set:
\begin{itemize}
    \item \textbf{Strict Accuracy (SA)}: Defined as whether the output contains the correct answer in the answer segment.
    \item \textbf{Accuracy}: Unlike SA, this metric considers the answer correct if the correct answer appears anywhere in the reasoning chain.
    \item \textbf{Chain-of-Thought Completeness (CTC)}: Evaluates whether the response includes a complete reasoning chain, assessing the model's ability to apply reasoning chains within limited lengths.
    \item \textbf{Answer Completeness (AC)}: Measures the completeness of the answer itself.
\end{itemize}

As demonstrated in Table \ref{tab:losses_revised}, the proposed QTHA achieves significantly higher training efficiency than classical LoRA under identical fine-tuning conditions for LLMs. When evaluated on chain-of-thought (CoT) reasoning tasks, QTHA exhibits statistically significant improvements in domain-specific metrics—including Contextual Task Consistency (CTC), Answer Coherence (AC), Accuracy, and Strict Accuracy—while reducing trainable parameters by over 76\% compared to LoRA.

Further testing on benchmark datasets (CPsyCounD, R1-Distill-SFT, and CH-R1-Math) confirms the robustness of QTHA. As summarized in Table \ref{tab:performance_metrics} and \ref{tab:ppl_metrics}, the framework demonstrates consistent improvements in text generation quality, with notable enhancements observed in the BLEU-4, ROUGE-1, ROUGE-2, ROUGE-L, PPL metrics.These results underscore QTHA's ability to reconcile parameter efficiency with enhanced model performance, establishing it as a viable alternative to conventional low-rank adaptation paradigms.

\begin{table}[!ht]
\centering
\caption{Cross-Architecture Performance Comparison by Dataset Scale.}
\label{tab:losses_revised}
\footnotesize
\begin{tabularx}{\columnwidth}{>{\centering\arraybackslash}X c c c c c c}
\toprule
\multicolumn{7}{c}{\textbf{CPsyCounD (Training/Validation Loss)}} \\
\cmidrule(lr){1-7}
\multirow{2}{*}{Dataset Scale} & \multicolumn{2}{c}{LoRA} & \multicolumn{2}{c}{QuanTA} & \multicolumn{2}{c}{QTHA} \\
\cmidrule(lr){2-3} \cmidrule(lr){4-5} \cmidrule(l){6-7}
 & Train & Val & Train & Val & Train & Val \\
\midrule
300 & 1.805 & 1.606 & 1.647 & 1.512 & \textbf{1.484} & \textbf{1.434} \\
600 & 1.657 & 1.514 & 1.565 & 1.478 & \textbf{1.452} & \textbf{1.421} \\
1000 & 1.577 & 1.447 & 1.520 & 1.437 & \textbf{1.440} & \textbf{1.405} \\
3000 & 1.764 & 1.600 & 1.633 & 1.517 & \textbf{1.572} & \textbf{1.494} \\

\midrule
\addlinespace[0.5em]
\multicolumn{7}{c}{\textbf{R1-Distill-SFT (Training/Validation Loss)}} \\
\cmidrule(lr){1-7}
\multirow{2}{*}{Dataset Scale} & \multicolumn{2}{c}{LoRA} & \multicolumn{2}{c}{QuanTA} & \multicolumn{2}{c}{QTHA} \\
\cmidrule(lr){2-3} \cmidrule(lr){4-5} \cmidrule(l){6-7}
 & Train & Val & Train & Val & Train & Val \\
\midrule
300 & 0.179 & 0.182 & 0.173 & 0.178 & \textbf{0.160} & \textbf{0.169} \\
600 & 0.172 & 0.162 & 0.170 & 0.163 & \textbf{0.159} & \textbf{0.158} \\
1000 & 0.169 & 0.160 & 0.168 & 0.163 & \textbf{0.160} & \textbf{0.159} \\
3000 & 0.229 & 0.213 & 0.220 & 0.208 & \textbf{0.211} & \textbf{0.202} \\

\midrule
\addlinespace[0.5em]
\multicolumn{7}{c}{\textbf{CH-R1-Math (Training/Validation Loss)}} \\
\cmidrule(lr){1-7}
\multirow{2}{*}{Dataset Scale} & \multicolumn{2}{c}{LoRA} & \multicolumn{2}{c}{QuanTA} & \multicolumn{2}{c}{QTHA} \\
\cmidrule(lr){2-3} \cmidrule(lr){4-5} \cmidrule(l){6-7}
 & Train & Val & Train & Val & Train & Val \\
\midrule
300 & 0.880 & 0.876 & 0.855 & 0.853 & \textbf{0.804} & \textbf{0.804} \\
600 & 0.829 & 0.788 & 0.821 & 0.791 & \textbf{0.776} & \textbf{0.767} \\
1000 & 0.800 & 0.765 & 0.804 & 0.781 & \textbf{0.773} & \textbf{0.763} \\
3000 & 0.852 & 0.819 & 0.830 & 0.803 & \textbf{0.789} & \textbf{0.762} \\

\bottomrule
\end{tabularx}

\vspace{0.5em}
\raggedright
\footnotesize
\textit{Note.} Parameter counts: LoRA = 1.26M (100\%), QuanTA = 0.73M (57.5\%), QTHA = 0.30M (23.7\%), additionally, the number of parameters trained based on R1-Distill-SFT is 0.57M(45.2\%). 
\end{table}

\begin{table}[ht]
\caption{Performance Metrics for CPsyCounD, R1-Distill-SFT and CH-R1-Math. Here, ``Simulator'' denotes the results of QTHA produced by a quantum circuit simulator; and ``Wukong'' produced by the superconducting quantum computer ``Origin Wukong''.}
\centering
\begin{tabularx}{\columnwidth}{>{\centering\arraybackslash}X c c c c}
    \toprule
    \multicolumn{5}{c}{\textbf{CPsyCounD}} \\
    \cmidrule(lr){1-5}
    Model & BLEU-4 & ROUGE-1 & ROUGE-2 & ROUGE-L \\
    \midrule
    LoRA & 14.823 & 39.560 & 17.216 & 34.239 \\
    QuanTA & 13.368 & 37.860 & 15.257 & 33.705 \\
    Simulator & 16.043 & 40.638 & 17.614 & 35.892 \\
    Wukong & \textbf{18.056} & \textbf{42.668} & \textbf{20.322} & \textbf{37.172} \\

    \toprule
     \multicolumn{5}{c}{\textbf{CH-R1-Math}} \\
    \cmidrule(lr){1-5}
    Model & CTC & AC & Accuracy & Strict Accuracy \\
    \midrule
    LoRA & 90 \% & 83 \% & 97 \% & 83 \% \\
    QuanTA & 97 \% & 93 \% & 97 \% & 93 \% \\
    Simulator & \textbf{100 \%} & \textbf{100 \%} & 93 \% & 93 \% \\
    Wukong & \textbf{100 \%} & \textbf{100 \%} & \textbf{100 \%} & \textbf{100 \%} \\
    \bottomrule
\end{tabularx}
\label{tab:performance_metrics}
\end{table}

\begin{table}[ht]
\caption{PPL for CPsyCounD, R1-Distill-SFT and CH-R1-Math. ``Simulator'' or ``Wukong'' means QTHA executed on a quantum circuit simulator or ``Origin Wukong'', respectively.}
\centering
\begin{tabularx}{\columnwidth}{>{\centering\arraybackslash}X c c c c}
    \toprule
   
    Model & PPL(CPsyCounD) & PPL(CH-R1-Math) & PPL(R1-Distill-SFT)  \\
    \midrule
    LoRA & 5.5785 & 2.5235 & 1.3355  \\
    QuanTA & 5.0758 & 2.4675 & 1.3306  \\
    Simulator & 4.8350 & 2.3556 & 1.3241  \\
    Wukong & \textbf{4.8325} & \textbf{2.3550} & \textbf{1.3237}  \\
    \bottomrule
\end{tabularx}
\label{tab:ppl_metrics}
\end{table}

\subsection{Implementation on quantum hardware}
QTHA demonstrates excellent performance on noisy quantum computers, while the output of LLMs is inherently a probability distribution matrix, its impact on the final result is always constrained within a specific threshold range, naturally exhibiting stochastic characteristics. The inherent random noise in quantum systems (such as decoherence and gate operation errors) may enhance reasoning robustness through the following mechanisms: in the probability matrix output by LLMs, only when the probability of a key token exceeds a preset threshold will it significantly influence decision-making. This implies that minor perturbations introduced by noise may be filtered out by these thresholds, thereby maintaining result stability. The probabilistic nature of quantum systems (e.g., superposition state collapse) and the probabilistic output of LLMs share mathematical similarities, and noise may implicitly calibrate the probability distribution to improve reasoning consistency.

We have constructed a variational quantum circuit module incorporating the Origin Quantum Cloud (QCloudService) in our model architecture. This module uses a quantum cloud service with a superconducting quantum computer backend named “Origin Wukong”, transforming user-defined variational quantum circuits (including input data and trainable parameters) into quantum circuit intermediate representations (OriginIR)~\cite{dou2022qpanda}. Given that the backend exclusively support Z-axis observable measurements, basis transformation operations are implemented to convert user-defined non-Z observables into equivalent Z-axis measurements. Specifically, each Pauli operator measurement generates a distinct quantum circuit.

Furthermore, to accommodate multi-dimensional input data (where input dimensions may exceed unity), each data dimension is entangled with trainable parameters in the variational quantum circuit to form a complete quantum architecture. Consequently, the total number of quantum circuits executed per batch on the quantum hardware is governed by the relationship:
\begin{equation}
    \mathrm{T = B \times Q},
\end{equation}
where B denotes the batch size and Q represents the number of qubits in the system.

To comply with quantum hardware batch processing constraints, we leverage the backend's asynchronous batch measurement interface run to partition quantum circuits into groups for sequential submission to QCloudService. As run operates asynchronously, we continuously monitor task status through the returned QCloudResult object until all measurement outcomes are retrieved.

Finally, measurement results are reformatted into machine learning-compatible tensors through concatenation and reshaping operations. These processed outputs can be directly processed by subsequent neural network modules, establishing seamless integration between quantum computations and classical machine learning pipelines.

\section{Conclusion}


This paper introduces QTHA, a quantum tensor hybrid adaptation framework for fine-tuning LLMs, achieving significant breakthroughs in parameter reduction and performance enhancement by integrating quantum tensor networks with quantum neural networks. Leveraging MPO decomposition, the framework transforms high-dimensional linear transformations into low-dimensional matrix product sequences via rank constraints, drastically reducing parameter counts while preserving tensor structures to effectively capture multi-dimensional data correlations. Additionally, the QNN generates high-dimensional nonlinear features unattainable by classical methods, and its fusion with the MPO network significantly enhances representation power in low-rank spaces. Experimental results demonstrate superior training loss reduction, even with limited datasets of 3,000 samples, highlighting quantum computing's potential to address computational bottlenecks in LLMs. Looking forward, the research will explore large-scale quantum pre-training schemes and fully quantum weight matrix reconstruction using variational quantum circuits to capture higher-order correlations. This work not only achieves lightweight optimization for billion-parameter models but also establishes a practical paradigm where LLM-driven tasks advance quantum hardware development. It lays the foundation for future quantum-enhanced Artificial General Intelligence systems, marking a key step toward scalable and engineerable quantum AI solutions.

\bibliographystyle{unsrt}
\bibliography{egbib}

\end{CJK*}

\end{document}